\documentclass[conference,compsoc]{IEEEtran}
\pdfoutput=1
\usepackage{booktabs}
\usepackage{amssymb}
\usepackage[cmex10]{amsmath}
\usepackage{listings}
\usepackage{graphicx}
\usepackage{url}
\usepackage{subfig}

\newcommand{\abs}[1]{\left\lvert#1\right\rvert}

\begin{document}
\title{Verificarlo: checking floating point accuracy
  through Monte Carlo Arithmetic}

\author{
  \IEEEauthorblockN{Christophe Denis\IEEEauthorrefmark{1}, Pablo de Oliveira
    Castro\IEEEauthorrefmark{2}, Eric Petit\IEEEauthorrefmark{2}}
  \IEEEauthorblockA{\IEEEauthorrefmark{1}CMLA - Centre de Math\'ematiques et de
    Leurs Applications, ENS Cachan\\
    Universit\'e Paris-Saclay et CNRS UMR 8536, 94235 Cachan France\\Email: christophe.denis@cmla.ens-cachan.fr}
  \IEEEauthorblockA{\IEEEauthorrefmark{2}Universit\'e de Versailles
    Saint-Quentin-en-Yvelines\\ Universit\'e Paris-Saclay\\Email: \{pablo.oliveira, eric.petit\}@uvsq.fr}
}

\maketitle

\begin{abstract} Numerical accuracy of floating point computation is a well
studied topic which has not made its way to the end-user in scientific computing.
Yet, it has become a critical issue with the recent requirements for code
modernization to harness new highly parallel hardware and perform higher
resolution computation. To democratize numerical accuracy analysis, it is
important to propose tools and methodologies to study large use cases in a
reliable and automatic way. In this paper, we propose verificarlo, an extension to
the LLVM compiler to automatically use Monte Carlo Arithmetic in a transparent way
for the end-user. It supports all the major languages including C, C++, and
Fortran. Unlike source-to-source approaches, our implementation captures the
influence of compiler optimizations on the numerical accuracy. We illustrate how
Monte Carlo Arithmetic using the verificarlo tool outperforms the existing
approaches on various use cases and is a step toward automatic numerical analysis.

\end{abstract}

\section{Introduction}

This paper presents a new compiler tool  to assess the uncertainties on a
scientific code due to the floating point (FP) arithmetic. It builds upon the extensive
work of Parker~\cite{PARKER1997} and Frechtling~\cite{frechtling2014tool} on Monte
Carlo Arithmetic (MCA) for floating point accuracy verification.  Floating point
computations are a model of real number computation where a real number is rounded
towards a floating point number, and some arithmetical properties, such as the
associativity of the sum, are lost. Consequently, the computer numerical results
are sensitive to the evaluation order of the floating point arithmetical
operations, the floating point precision, and the rounding mode.

The quantification of the floating point uncertainties is important. In the
next decade, exascale supercomputers will provide the computational power required
to perform very large scale simulations. For certain applications the results of
exascale simulations will be of such high resolution that experimental
measurements will be insufficient for validation purposes. As floating
point approximations of numeric expressions are neither associative nor
distributive, the results of a numerical simulation can differ between
executions. As reported by Duff~\cite{DUFF}, ``{\it Getting different results
for different runs of the same computation can be disconcerting for users even if,
in a sense, both results are correct}''. There is a need to have an automatic and
global approach giving a confidence interval on the results taking into account
the floating point arithmetic effect.

Numerical verification is a procedure to estimate the effect of the floating
point model on the accuracy of the computed results. It is the first step of a
rigorous Verification and Validation (V\&V) procedure. Several
methods exist to perform a numerical verification on a numerical
code. Kahan, the primary architect of
the IEEE-754 standard for floating point computation, argues in
\cite{kahan2006futile} that using extendable precision interval arithmetic is
almost foolproof. The interval arithmetic is an arithmetic defined on sets of
guaranteed intervals rather than on sets of IEEE-754 numbers. The numerical
verification on a scientific code using IEEE-754 floating point numbers consists of
comparing the results with those obtained on a shadow code using interval
arithmetic. It requires that the results intervals are sufficiently small. If not,
the computation needs to be performed again by extending the precision of the
interval arithmetic. Unfortunately, even if it guarantees the result, interval
arithmetic typically produces overly pessimistic bounds as it does not take into
account the round-off error compensation when using the rounding mode to the
nearest. Some numerical algorithms need to be adapted when using interval
arithmetic. For example, the Newton-Raphson method needs to be modified in order to
obtain convergence under interval
arithmetic~\cite{revol2003interval}. Consequently, from an industrial point of
view, it is only possible to use extendable precision interval arithmetic on
specific numerical algorithms and not on a whole scientific code.

An alternative is to compute stochastic confidence intervals on the results by
applying random perturbations on the numerical operations. For example, the Discrete
Stochastic Arithmetic implemented in the CADNA library perturbates computations by
randomly changing the rounding mode. Discrete Stochastic Arithmetic is based on
CESTAC developed by Vignes in 1974.  CADNA is a powerful numerical debugger tool
which has been used to solve real problems. Nevertheless, CADNA has some
limitations. First, Chatelin and Parker~\cite{chatelin88,PARKER1997} show that the
CESTAC assumptions are not always verified when performing numerical analysis,
which can introduce errors in CADNA estimations. Second, using CADNA requires
manually modifying the original program sources to use special CADNA types. For
large code bases, this process can be costly.

In this paper, we propose the following contributions:
\begin{itemize}
\item Verificarlo, a new LLVM compiler tool to automatically use the Monte Carlo
arithmetic in place of the IEEE-754 FP. Verificarlo operation is transparent for
the user and does not require manually modifying the source code.

\item A set of experiments to validate the automatic MCA approach using
verificarlo and compare it to the state-of-the-art MCA approach using
CIL~\cite{frechtling2014tool} and CESTAC approach using CADNA~\cite{VIGNES2004}.

\end{itemize}

This paper is organized as follows.  Section~\ref{sec:pbmeth} presents stochastic
arithmetic for numerical verification.  Section~\ref{sec:verif} introduces the
verificarlo tool, its advantages and limitations, and compares it to other
approaches.  Finally, section~\ref{sec:xp} proposes a set of experiments to
validate verificarlo and demonstrate its capabilities.

\section{Probabilistic methods to check the floating point accuracy}
\label{sec:pbmeth}
The aim of this section is to briefly present two probabilistic methods used to
check the floating point accuracy:
the Monte Carlo Arithmetic (MCA) and the Discrete Stochastic Arithmetic (DSA).

\subsection{Monte Carlo Arithmetic (MCA)}
\label{sec:mca}

Monte Carlo Arithmetic (MCA) tracks rounding and catastrophic cancellation errors
at a given virtual precision $t$ by applying randomization to input and output
operands. MCA makes no assumption about the round-off error distribution and
produces unbiased random round-off errors.

It forces the results of floating point operations to behave
like random variables. This turns executions into trials of a Monte Carlo
simulation allowing statistics on the effects of rounding error to be obtained
over a number of executions. This section succintly summarizes MCA, for a full
presentation see~\cite{PARKER1997}.

To model errors on a FP value $x$ at virtual precision $t$,
Parker proposes the following function:
\begin{equation}
  inexact(x) = x + 2^{e_x-t}\xi,
\end{equation}
where $e_x$ is the exponent of the FP value $x$ defined as $e_x = \lfloor \log_2{|x|} + 1 \rfloor$
and $\xi$ is a uniformly distributed random variable in the open interval $(-\frac{1}{2},\frac{1}{2})$.

Each floating point operation $x \circ y$  is transformed into a MCA FP operation
using one of the following modes:
\begin{itemize}
\item RR: Random Rounding, which tracks rounding errors by introducing an
error in the outbound value,

\begin{equation*}
       x \circ y  \rightarrow round(inexact(x\circ y))
\end{equation*}

\item PB: Precision Bounding, which tracks catastrophic cancellations by
introducing errors in the inbound values,
\begin{equation*}
       x \circ y  \rightarrow round(inexact(x)\circ inexact(y))
\end{equation*}

\item Full MCA: Monte Carlo Arithmetic with inbound and outbound errors,
\begin{equation*}
       x \circ y  \rightarrow round(inexact(inexact(x)\circ inexact(y)))
\end{equation*}
\end{itemize}

When the exact solution $x$ of a problem is known, we can measure the number of
significant digits $s$ in base $\beta$ by computing the magnitude of the relative
error between the approximated value $\hat x$ and the exact value $x$ using the
following formula
\begin{equation}
\label{eq:ra}
s = -\log_{\beta}{\abs{\frac{\hat x-x}{x}}}
\end{equation}

Parker extends this definition to MCA and shows~\cite[p.~23]{PARKER1997} that the
total significant digits for a set of MCA results at virtual precision $t$ is
given by the magnitude of the relative standard deviation
\begin{equation}
\label{eq:rmca}
s' = -\log_{\beta}{\frac{\sigma}{|\mu|}}
\end{equation}
In this formula, $\mu$ and $\sigma$ are the mean and the standard deviation of the
result distribution. Unfortunately, the exact distribution of results is unknown,
but it can be empirically estimated by using a large number of Monte Carlo
trials. Indeed for a large number of trials,
$s' \approx -\log_{\beta}{\frac{\hat \sigma}{|\hat \mu|}}$, where
$\hat \mu$ and $\hat \sigma$ are the sample mean and sample
standard deviation.

The metrics given by equations~\ref{eq:ra} and ~\ref{eq:rmca} will be used in
section~\ref{sec:xp} to evaluate our outputs and compare to other approaches.

\subsection{Discrete Stochastic Arithmetic (DSA)}

Discrete Stochastic Arithmetic (DSA) is based on the CESTAC method. The CESTAC
method is a pioneer work in the domain of the random computer
arithmetic~\cite{VIGNES1974}. The ingenious idea is to randomly change the
rounding mode of a floating point (FP) computation to estimate its accuracy. For
debugging purposes, DSA has made the choice to carry out a single program in which
each FP operation is performed $N$ times with a rounding mode towards plus or
minus infinity. For each sample, the rounding mode is randomly chosen. There is
thus a probability $P_{N}=2^{1-N}$ that all the $N$ samples compute a FP operation
with the same rounding mode. The number of significant digits is computed by using
the Student distribution.  DSA also redefines relational operators. A full review
of DSA is provided in~\cite{VIGNES2004}.

The CADNA library is an implementation of Discrete Stochastic Arithmetic (DSA)
with $N=3$ samples. The first two samples compute each FP operation with a random
rounding mode whereas the last one uses the rounding mode not used by the second
sample.

The validation of CESTAC method and DSA is based on a probabilistic first
order model. It has been established by considering that elementary round-off
errors of the FP arithmetic operations are random independent, centered and
uniformly distributed variables. Kahan has formulated strong objections to this
assumption by proposing in~\cite{kahan1996improbability} the following case study:

\begin{equation}
\label{eq1}
Kh_x(dx)=cf(x+dx)-rp(x)
\end{equation}
\begin{equation}
cf(x)=4-\frac{3(x-2.0)*((x-5.0)^2+4)}{x+(x-2.0)^2((x-5.0)^2+3.0)}
\end{equation}
\begin{equation}
rp(x)=\frac{622.0-x(751.0-x(324-x(59.0-4.0x)))}{112-x(151.0-x(72.0-x(14.0-x)))}
\end{equation}
with :
\begin{itemize}
\item $x= 1.60631924$
\item $dx=i\epsilon, \mbox{i varying from 1 to 300 with step 1}, \epsilon=2^{-53}$
\end{itemize}

\begin{figure*}
   \begin{minipage}[c]{.30\linewidth}
      \includegraphics[scale=0.54]{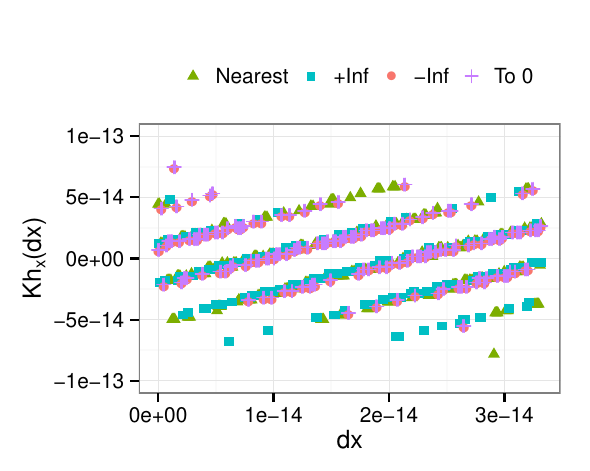}
      \caption{Eq.~\ref{eq1} using IEEE-754 DP \label{ieee}}
   \end{minipage} \hfill %
   \begin{minipage}[c]{.30\linewidth}
      \includegraphics[scale=0.54]{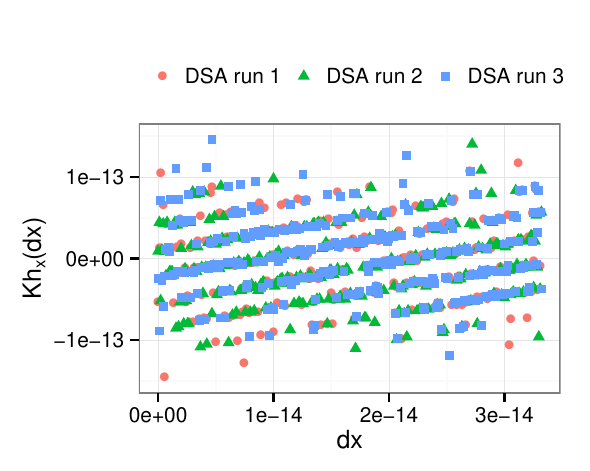}
       \caption{Eq.~\ref{eq1} using CADNA \label{dsa}}
   \end{minipage} \hfill %
   \begin{minipage}[c]{.30\linewidth}
      \includegraphics[scale=0.54]{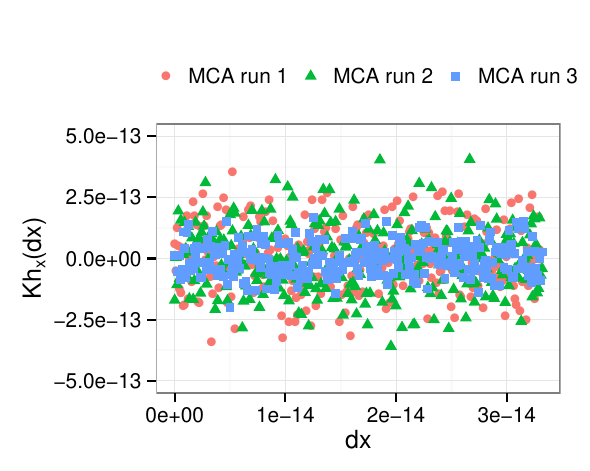}
       \caption{Eq.~\ref{eq1} using verificarlo \label{mca}}
   \end{minipage}
\end{figure*}

Figure~\ref{ieee} presents the evaluation of equation~\ref{eq1} by using IEEE-754
double precision floating point numbers. On one hand, the stripped patterns in
figures~\ref{ieee} and~\ref{dsa} show that IEEE-754 and CADNA results are not
uniformly distributed random variables.  On the other hand, the Monte Carlo
Arithmetic introduced in section~\ref{sec:mca} permits to obtain an independent
and identically distributed (iid) centered uniform sample as shown in
figure~\ref{mca}.

Chesneaux and Vignes argue in \cite{chesneauxvignes} that even if the independent,
centered, and uniform assumption is not satisfied, the CESTAC method is able to
correctly estimate the number of significant digits with a probability of $95\%$.
Nevertheless, Chatelin~\cite{chatelin88} indicates that CESTAC's {\it confidence
cannot be greater than 5\% under various conditions which are shown to be often
met in practice} and Parker~\cite{PARKER1997} explains how MCA can overcome these
limitations.

It is important to notice that DSA and MCA also differ from a methodological point
of view. DSA uses a synchronous approach where the user incrementally fixes the
numerical instabilities reported by CADNA. DSA is based on a first order model:
some operations such as unstable division or unstable multiplication may invalidate
the model. The user must correct these unstable FP operations before CADNA can
estimate accurately the number of significant digits. Therefore DSA is well suited
to perform numerical debugging to correct numerical instabilities as done
in~\cite{SCOTT2007}.

\section{Verificarlo: A software for automatic Monte Carlo Arithmetic analysis}
\label{sec:verif}

As previously discussed in section~\ref{sec:mca}, Monte Carlo Arithmetic is a powerful framework to understand the numerical
stability of a function or program. To encourage its wide adoption by the
community we have developed verificarlo, a tool for automatic MCA analysis of C,
C++ and Fortran programs. Verificarlo builds upon the LLVM
Compiler~\cite{lattner2004llvm} project and the
MCALIB~\cite{frechtling2014tool}. It takes as input a source code project and
compiles it with a special instrumentation pass that replaces all floating point
operations by their MCA counterpart in MCALIB. The instrumentation can be applied
to the whole program or only to a function of interest.

Two previous approaches for automatic MCA simulation have been proposed.  Yeung et
al.~\cite{yeung2011monte} implement MCA at the hardware level through specialized
FPGA co-processors. While providing low overhead, this approach is impractical
because it requires specialised hardware not available to the practitioner.

Frechtling et al.~\cite{frechtling2014tool} leverage source-to-source rewriting of
floating point operations through the CIL tool for program transformation~\cite{necula2002cil}. The first
drawback of using CIL is that analysis is limited to C programs. The second and
main drawback with source-to-source rewriting is that the instrumentation happens
before and may hinder the compiler optimizations. That means that the floating
point operations in the MCA binary and in the original binary may be
different. What is tested is not always what will be executed because CIL cannot
capture the effect of compiler optimizations on numerical errors.

To tackle these issues, verificarlo instruments the floating point at the
optimized Intermediate Representation level (IR). First, because the IR
representation is independent of the source language used, verificarlo can operate
on any source language supported by LLVM that includes C and C++ through clang and
Fortran through dragonegg. Second, the instrumentation pass is done after all the
other front-end and middle-end optimization passes (which include all the floating
point optimizations such as \texttt{-ffast-math} or \texttt{-freciprocal-math}).

Verificarlo is available at \url{http://www.github.com/verificarlo/verificarlo} under an
open source licence. Verificarlo computes Monte Carlo arithmetic using a modified
version of MCALIB. One notable difference is that our version of MCALIB replaces
the standard libc pseudo-random generator with Mersenne
Twister~\cite{matsumoto1998mersenne}. This provides two benefits: first for user
programs using the libc \texttt{rand} function, having a separate generator avoids
seeding collisions. Second, Mersenne Twister is a robust random number generation
in the context of Monte Carlo simulations~\cite{matsumoto1998dynamic}.

One disadvantage of MCA is that it requires a large number of samples compared to
DSA and is therefore more costly. Table~\ref{tab:perf} compares the cost of
running a numerical analysis with CADNA, MCALIB and verificarlo.  Verificarlo and
MCALIB are significantly slower than CADNA. The first reason is that to be
accurate they require a higher number of samples. The second reason is that both
MCALIB and verificarlo use the MPFR~\cite{fousse2007mpfr} library to compute MCA
samples. Performing high precision computations with MPFR is more costly than
changing the rounding mode.

Fortunately, verificarlo supports massively parallel execution out of the box. The
high overhead can be mitigated by concurrently measuring the MCA samples. Our
tests show an ideal scalability thanks to the embarrassingly parallel nature of
Monte Carlo. In contrast, CADNA parallelization does not
scale~\cite{jezequel2013parallelization} because it requires explicit synchronization
between processes.

\begin{table}
 \begin{tabular}{lrrr}
    \toprule
    version  & samples & total time (s) & time/sample (s) \\
    \midrule
    original program     & 1       & .056           & .056 \\
    CADNA                & 3       & 2.93           & .097 \\
    MCALIB               & 128     & 1184.02        & 9.25 \\
    verificarlo          & 128     & 834.57         & 6.52 \\
    verificarlo 16 threads & 128     & 54.39         & .42  \\
    \bottomrule
  \end{tabular}
  \caption{Performance in seconds for the numerical analysis of the compensated
sum algorithm detailed in section~\ref{sec:sc1} on an array of size 1000000. All
the binaries were compiled using \texttt{-O0}. The experiment was performed on a
16-core 2-socket Xeon E5@2.70GHz with 20Mb L3 cache per socket and 64Gb of
RAM. \label{tab:perf}}
\end{table}

\section{Experimental results}
\label{sec:xp}

This section presents four case studies to illustrate floating point accuracy
verification using verificarlo and its benefits compared to other state-of-the-art
approaches: CADNA C 1.1.9~\cite{lamotte2010cadna_c} and CIL+MCALIB.

The first case study evaluates the numerical error in a compensated sum algorithm
using CADNA, CIL+MCALIB, and verificarlo. Among the three tools, only verificarlo
is able to capture the effect of compiler flags on numerical errors.

The second case study deals with the solving of a linear system $Ax=b$ proposed by
Kahan in~\cite{KAHAN66}. The matrix $A$ is ill-conditioned which can reduce the
number of significant digits. This case study demonstrates how verificarlo using
MCA can estimate the number of significant digits.  The resolution is done by using
the sophisticated LAPACK routines.  It has not been possible to use CADNA as it
requires to manually change the source code of the LAPACK library.
Verificarlo gives an estimation of the number of significant digits close to the
number of significant  digits between the IEEE-754 computing and the exact solution.

The third case study deals with unstable branching. In this case,
CADNA is too pessimistic as it estimates that the numerical result has no accurate
digits whereas verificarlo finds a number of significant digits close to the number
of accurate significant digits between the IEEE-754 computing and the exact value.

The fourth case study deals with the management of a counter. The comparison between
the IEEE-754 double precision computing and the exact solution shows that the
numerical result is a numerical noise having no significant digits.
Unfortunately for this case, CADNA is too optimistic as it over-estimates the
result significant digits whereas verificarlo succeeds to estimate that the
IEEE-754 DP result is a numerical noise.

\subsection{Case study 1: Compensated Summation}
\label{sec:sc1}

\begin{figure}

\lstset{language=C,frame=lines,numbers=left,numberstyle=\footnotesize, basicstyle=\ttfamily\footnotesize, xleftmargin=.03\textwidth}
\begin{lstlisting}
float sum = f[0];
float c = 0.0, y, t;

for (int i=1;i<n;i++) {
    y = f[i] - c;
    t = sum + y;
    c = (t - sum) - y;
    sum = t;
}

return sum;
\end{lstlisting}

\caption{\emph{Kahan compensated summation:} with \texttt{-O3 -ffast-math} flags
the compiler simplifies and removes the computation of the compensation term
\texttt{c}. \label{lst:kahan}}

\end{figure}

  In the following, we demonstrate the importance of capturing compiler effects on a
standard use case: Kahan's compensated summation algorithm
\cite[p.~83]{higham2002accuracy} shown on figure~\ref{lst:kahan}.
The C implementation is particularly sensible to compiler optimizations when
floating point associativity rules are relaxed with \texttt{-ffast-math -O3}. The
compiler uses simple common subexpression elimination and rewrites line 9 as
\verb! sum = sum + f[i]! which is the naive non-compensated summation.

Using verificarlo and CIL+MCALIB~\cite{frechtling2014tool} we measured 1000 sample
executions of the Kahan summation code compiled with \texttt{-O3 -ffastmath} and
\texttt{-O0}. The input array contains 1000 random single precision floats in the
interval $[0,1]$ and therefore has a condition number of 1.  Only Random Rounding
MCA errors were considered in this study.

\begin{figure}
  \includegraphics[width=\linewidth]{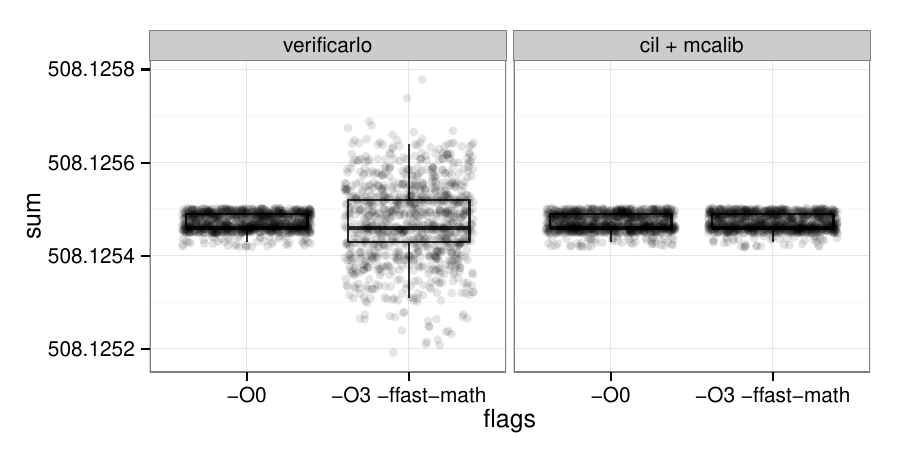}
\caption{
\emph{One thousand MCA RR samples of Kahan summation:} CIL+MCALIB is unable to capture the compiler effect on Kahan's summation because it
operates at the source level. On the other hand verificarlo operates after
compiler optimizations and correctly shows that the \texttt{-O0} version is more
precise than the \texttt{-O3 -ffast-math} version thanks to the compensation term
\texttt{c}. \label{fig:kahan}}
\end{figure}

\begin{table}
  \centering
\begin{tabular}{lr@{\extracolsep{.5cm}}r}
  \toprule
                          & -O0 & -O3 -ffast-math\\
  \midrule
  CADNA                   & 7   & 7 \\
  CIL+MCALIB              & 7.3 & 7.3 \\
  verificarlo             & 7.3 & \textbf{5.8} \\
  \bottomrule
\end{tabular}
\caption{Number of significant digits estimated with size $n=100000$. Verificarlo is
  the only tool that detects that \texttt{-O3 -ffast-math} introduces a loss of accuracy.
 \label{tab:kahan_digits}}
\end{table}

\begin{figure}
  \centering
  \includegraphics[width=.7\linewidth]{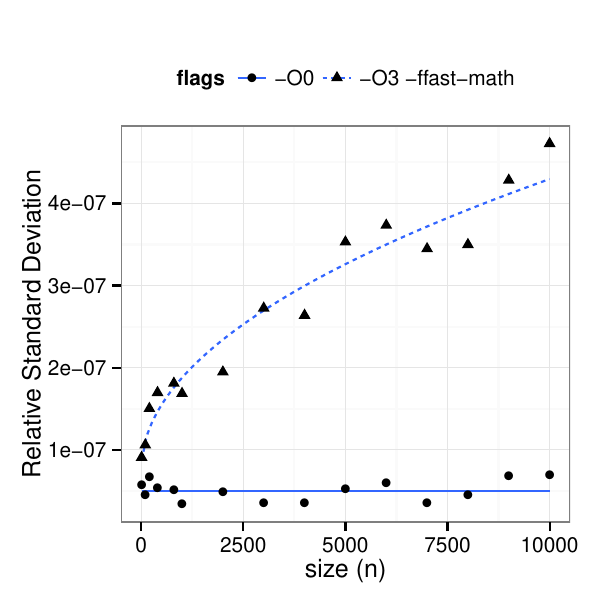}
  \caption{Relative Standard Deviation ($\frac{\sigma}{|\mu|}$) of Kahan's
    compensated sum computed on 1000 verificarlo samples.  The compensated
    \texttt{-OO} version has a constant error while the \texttt{-O3 -fast-math}
    error increases as $O(\epsilon\sqrt{n})$. \label{fig:kahansize}}
\end{figure}

Figure~\ref{fig:kahan} compares the results between verificarlo and CIL+MCALIB.
On one hand, CIL+MCALIB is unable to detect any difference between the two
versions. Table~\ref{tab:kahan_digits} shows the number of significant digits
predicted by CADNA, CIL+MCALIB and verificarlo for an array of 100000
floats. Again, CADNA and CIL+MCALIB are blind to compiler optimizations because
they operate at source level. On the other hand, verificarlo correctly shows the
loss of accuracy in the \texttt{-O3 -ffast-math} version.

In figure~\ref{fig:kahansize} we plot the relative standard deviation of
verificarlo's samples with different input sizes. Theoretical error analysis~\cite[p.~85]{higham2002accuracy} shows
that Kahan's compensated sum relative error is bounded by $2\epsilon +
O(n\epsilon^2)$ where $n$ is the input size and $\epsilon$ the computation's
precision. So Kahan's sum relative error is constant for inputs satisfying
$n\epsilon < 1$. However the relative error of a naive sum grows as
$O(\epsilon\sqrt{n})$ when floating point errors are iid with zero mean.  We see
that verificarlo stochastic error analysis closely matches the theoretical error
bounds.

This experiment demonstrates how the late instrumentation in verificarlo helps
evaluating the impact of compiler optimizations on numerical stability.

\subsection{Case study 2:  Resolution of a linear system}
Kahan~\cite{KAHAN66} proposes the following linear system with a large condition number:

\begin{equation}
\label{eq::kahan_lin_system}
\left( \begin{array}{cc}
0.2161& 0.1441  \\
1.2969 & 0.8648  \\
\end{array} \right) x =
\left( \begin{array}{c}
0.1440   \\
0.8642   \\
\end{array} \right)
\end{equation}
The exact solution of equation~\ref{eq::kahan_lin_system} is:
\begin{equation}
 x =
\left( \begin{array}{c}
2   \\
-2   \\
\end{array} \right)
\end{equation}

In the context of this case study, we solve equation~\ref{eq::kahan_lin_system}
using the LAPACK numerical library. LAPACK is written in Fortran 90 and provides
sophisticated routines for solving systems of simultaneous linear equations,
least-squares solutions of linear systems of equations, eigenvalue problems, and
singular value problems.
Table~\ref{linear_ieee} reports the results of the resolution using the
IEEE-754 single precision and double precision arithmetic with a rounding mode to
the nearest.

Unfortunately, it has not been possible to use CADNA as it
requires to modify the source code which is difficult and costly to do in a whole
numerical library such as LAPACK. For example, Montan~\cite{montan} has developed
a modified version of the LAPACK DGEMM routine (matrix multiplication) to
efficiently use CADNA.  In contrast, the use of verificarlo has permitted to
implement automatically MCA on the whole LAPACK library.  The number of samples
used by MCA in this experiment is set to 1000. Table~\ref{linear_mca} reports the
results of the resolution by using the MCA single and double precision
floating point arithmetic.

\begin{table}
\begin{tabular}{l@{\extracolsep{.25cm}}rr}
  \toprule
FP arithmetic & Result &  s \\
  \midrule
IEEE-754 single precision & $x(1) =  1.33317912$          &  $0$  \\
(default rounding)        & $x(2) = -1.00000000$          &  $0$  \\
  \midrule
IEEE-754 double precision & $x(1) = 2.00000000240030218$   &  $9$  \\
(default rounding)        & $x(2) = 2.00000000359962060$   &  $9$  \\
  \bottomrule
\end{tabular}
\caption{Resolution of equation~\ref{eq::kahan_lin_system} by using the  IEEE-754 single and double precision floating point arithmetic and comparison to the exact solution}
\label{linear_ieee}
\end{table}

\begin{table*}
\begin{center}
\begin{tabular}{l@{\extracolsep{.25cm}}llr}
  \toprule
FP arithmetic & Mean value & Standard deviation & s' \\
  \midrule
MCA single precision & $\bar{x}(1)= 1.02463705$   & $\sigma(x(1))= 6.46717332$  &  $0.0$ \\
                     & $\bar{x}(2)= 6.46717332$   & $\sigma(x(2))= 9.69851698$  &  $0.0$  \\
MCA double precision & $\bar{x}(1)= 1.9999999992$  & $\sigma(x(1))= 8.4541287415\times10^{-9}$  &  $8.3$ \\
                     & $\bar{x}(2)=-1.9999999988$ &  $\sigma(x(2))= 1.26782603316\times10^{-8}$ &  $8.2$ \\
  \bottomrule
\end{tabular}
\end{center}
\caption{Resolution of equation~\ref{eq::kahan_lin_system} by using the MCA  single and double precision floating point arithmetic}
 \label{linear_mca}
\end{table*}

In this example, the estimator given in section~\ref{sec:mca} accurately computes the
number of significant digits. Moreover, this estimation does not require knowing
beforehand the exact solution of the system.

\subsection{Case study 3: Unstable branching}

This section presents the differences between MCA and DSA when dealing with
branches testing a FP value. The following C program is used in this case study:

\lstset{language=C,frame=lines,numbers=left,numberstyle=\footnotesize, basicstyle=\ttfamily\footnotesize, xleftmargin=.03\textwidth}
\begin{lstlisting}
  double a,b,c;
  a=2.0*sqrt(3.0)/3.0;
  b=a*a-a*a;
  if (b>=0)
    c=sqrt(b)+10.0;
  else
    c=sqrt(-b)+10.0;
  return c;
\end{lstlisting}

The test on $b$ at line 4 prevents the square root computation of a negative
number. In the IEEE-754 standard, the square root of a negative number returns NaN
(Not A Number).

Table~\ref{d_all} compares the exact value of $d$, its numerical evaluation by
using IEEE-754 double precision with the rounding mode to the nearest and the
numerical verification done both by CADNA and verificarlo.
\begin{table*}

\begin{center}
\begin{tabular}{l@{\extracolsep{.25cm}}l@{\extracolsep{.25cm}}l@{\extracolsep{.25cm}}}
  \toprule
FP arithmetic & Result &        Significant digits\\
  \midrule
Exact solution            &   $d=10.$                  &   \\
IEEE-754 double precision &   $d=10.$                  &  $s=+\infty^\dagger$\\
  \midrule
CADNA double precision    &   $d =@.0 $                &  $0$ \\
  \midrule
verificarlo               &   $\bar d = 10.$      &  $s'=9.13$    \\
  100 samples             &   $\sigma(d)=7.255009\times10^{-09}$ &        \\
\bottomrule

\end{tabular}

\caption{Comparison between the exact value of $d$, its numerical evaluation by
using IEEE-754 double precision with the rounding mode to the nearest and the
numerical verification done both by CADNA and verificarlo. $@.0 $ means that CADNA
found that the result has no significant digits.\\
($\dagger$) $+\infty$ corresponds to the maximum number of representable
significant digits, in the case of a IEEE-754 double between 15 and 17.
  \label{d_all}
}
\end{center}
\end{table*}

The estimation of the number of significant digits (9.13 decimal digits) is
reasonable given that the rounded IEEE-754 computation is exact.  On the other
hand, the numerical verification performed by CADNA indicates that the result $c$
is a numerical noise having no significant digit.

In verificarlo, each sample execution can follow a different branch in
the code; executions are independent. CADNA, unlike verificarlo, works in a synchronous mode. Each floating
point operation is computed three times with different rounding mode towards plus
or minus infinity. Then a reconciliation process is used to select a single branch
outcome for the three CESTAC traces.  For the test at line 4, the three
samples of $b$ are $b_1=-0$, $b_2=2.22045\times10^{-16}$ and
$b_3=-2.22044\times10^{-16}$. In this case, CADNA reconciliation uses the mean of
the three traces, which is positive, and concludes that the test is
true. Unfortunately, the square root evaluation on the third negative
sample produces an invalid NaN value. This case study shows that CADNA can produce
invalid results on branch programs. Using CADNA on large code bases
requires the help of an expert to detect invalid results due to branching.
In contrast, MCA uses independent computing on these samples so no invalid
computation is performed during this case study.

\subsection{Case study 4: Alternating counter}

In the C code below, a counter $c$ is initialized to $5\times10^{13}$. The counter
is updated iteratively $N$ times, with $N = 10^{8}$. For each update, $c$ is incremented or
decremented by $10^{-6}$ depending on the parity of the iteration number.

\lstset{language=C,frame=lines,numbers=left,numberstyle=\footnotesize, basicstyle=\ttfamily\footnotesize, xleftmargin=.03\textwidth}
\begin{lstlisting}
  unsigned int i;
  double c=-5e13;
  for (i=0;i<100000000;i++) {
    if (i%2==0)
      c=c+1.e6;
    else
      c=c-1.e-6;
  }
  return c;
\end{lstlisting}

Assuming infinite precision, at the end of this process the expected exact value of $c$ should be:
\begin{equation}
c= -5\times 10^{13} + \frac{1}{2} 10^{8} 10^{6} -\frac{1}{2} 10^{8} 10^{-6}= -50
\end{equation}

Table~\ref{c_ieee} compares the exact value of $c$ to its numerical evaluation
when using IEEE-754 double precision with the rounding mode to the nearest, towards
$-\infty$ and towards $+\infty$.

\begin{table}

  \begin{tabular}{l@{\extracolsep{.5cm}}lr}
    \toprule
    FP arithmetic & Result & s \\
    \midrule
Exact solution          &   $c=-50.0$                             &  -            \\
    \midrule
    IEEE-754 double precision & & \\
    \hspace*{.25cm}rounded to the nearest&   $c=-0.02460...$              &  $0$           \\
    \hspace*{.25cm}rounded towards $-\infty$ &   $c=-2073773.08...$              &  $0$           \\
    \hspace*{.25cm}rounded towards $+\infty$ &   $c=-0.008202...$              &  $0$           \\
    \hspace*{.25cm}rounded towards $0$              &   $c=-0.008202...$           &  $0$           \\
    \bottomrule
\end{tabular}

\caption{Comparison between the exact value of $c$ and its numerical evaluation by
using IEEE-754 double precision with the rounding mode to the nearest, towards
$-\infty$ and towards $+\infty$}

  \label{c_ieee}
\end{table}

\begin{figure}
  \centering
  \includegraphics[width=.8\linewidth]{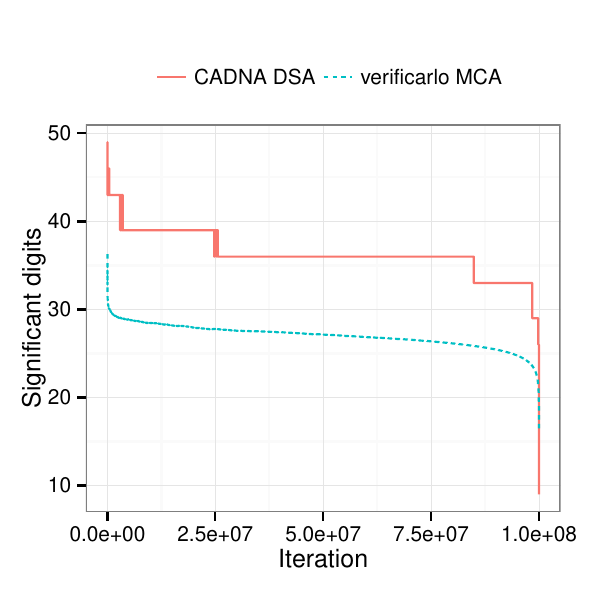}
  \caption{Evolution of the estimated significant number of bits with the iteration number}
  \label{fig:sigbit}
\end{figure}

The IEEE-754 arithmetic provides results having no significant digits whatever the
rounding mode used: there is then a strong numerical
problem. Table~\ref{c_cadna_mca} reports the results of the numerical verification
performed by CADNA and verificarlo and figure~\ref{fig:sigbit} the evolution of
the number of significant bits in the result estimated by verficarlo and CADNA.

\begin{table*}
\begin{center}
\begin{tabular}{l@{\extracolsep{.5cm}}l@{\extracolsep{.5cm}}l@{\extracolsep{.5cm}}}
  \toprule
FP arithmetic & Result &  Significant digits \\
  \midrule
CADNA DP &   $c=-0.103\times10^6$ &               3               \\
&   \multicolumn{2}{l}{CADNA detects no instabilities, only estimated significant digits are printed} \\
  \midrule
verificarlo  DP   &   $\bar c=-45.101042$ &               $s'=0$  \\
(1000 samples ) & $\sigma(c) = 44.33$                       &     \\
  \bottomrule
\end{tabular}
\caption{Numerical verification of the computation of $c$ performed both with CADNA and MCA}
\label{c_cadna_mca}
\end{center}
\end{table*}

CADNA in this case study overestimates the number of significant digits of the
result. Indeed, the result is a numerical noise with no significant
digit. Furthermore, CADNA overestimates the number of significant for all the
program iterations. The overestimation is not due to the small number of samples
(N=3) used by CADNA. The issue here is that each CADNA sample performs the
arithmetical operations with a rounding mode towards $+\infty$ or $-\infty$ with a
probability equal to $\frac{1}{2}$. As this case study is a linear problem, the
expected CADNA result should be the mean value of the $c$ values computed with
IEEE-754 rounding to $+\infty$ or $-\infty$ that is to say:

\begin{eqnarray*}
  E(c_{cadna})\simeq& \frac{-2073773.08...- 0.008202...}{2} \\
             \simeq& - 103686 \simeq  -0.103\times10^6
\end{eqnarray*}

In contrast, the numerical verification performed with verificarlo shows that
standard deviation is greater than the mean value of the MCA samples. It correctly
indicates that the result computed using IEEE-754 floating point numbers is a
numerical noise.

\section{Limitations and future work}

Verificarlo is a  fully automatic tool to instrument an application for numerical precision
analysis. The current version is stable and has been successfully used to analyse
small and large code bases, yet it is limited in some respects.

As shown in section~\ref{sec:verif}, the verificarlo runtime overhead is
high. This is due to MCA inexact computations being performed with MPFR. When the
desired virtual precision is low and known in advance, the overhead can be
reduced by performing computations using a fixed precision implementation
(e.g. double, quads) and avoiding the MPFR abstraction. This improvement is
scheduled for the next version of verificarlo.

Unlike CADNA, verificarlo does not support numerical debugging out of the box. In
the future we would like to include a mode that allows pinpointing the exact
operation or routine that is to blame for a precision loss.  We would also like to
include a statistical post-treatment toolbox to go beyond the standard deviation
analysis. This toolbox could help non-experts understand and interpret the output
of the MCA analysis.

Finally, it is important to test the robustness of the MCA approach on different
classes of numerical algorithms such as linear algebra or compensated algorithms and
also full-scale real-life applications.

\section{Conclusion}
The control of the numerical accuracy of scientific codes becomes crucial in
particular when using HPC ressources.  It is also necessary to control the
floating point computation when porting a scientific code on another programing
language or on different computing ressources. These tasks raise the need for a
tool that automatically estimates, without the assistance of an expert, the
interval of confidence of computed results.
Verificarlo is the first step toward a fully automatic tool. It is based on the
Monte-Carlo Arithmetic and uses a compiler approach to easily instrument the code
that is executed.

Our case studies illustrate the advantages of using verificarlo for numerical
analysis on scientific codes. They show that verificarlo overcomes some
methodological and technical limitations of the CADNA library to estimate the
numerical accuracy. Verificarlo is the first tool to implement MCA arithmetic at
the intermediate representation level. Unlike CADNA or MCALIB+CIL, this allows to
capture the effect of compiler optimizations on numerical accuracy.

Verificarlo is available at \url{http://www.github.com/verificarlo/verificarlo}
under an open source licence.

\bibliographystyle{alpha}
\bibliography{verificarlo-preprint.bib}

\end{document}